\documentclass[default,iicol]{sn-jnl}% Default with double column layout

\jyear{2021}%

\theoremstyle{thmstyleone}%
\theoremstyle{thmstyletwo}%

\theoremstyle{thmstylethree}%

\usepackage{bm}
\usepackage{accents}

\newcommand{\pt}{\partial}
\newcommand{\bnabla}{\bm{\nabla}}
\newcommand{\utilde}[1]{\undertilde{#1}}

\begin{document}

\title[Article Title]{Generative Adversarial Networks to infer velocity components in rotating turbulent flows}


\author*[1]{\fnm{Tianyi} \sur{Li}}\email{tianyi.li@roma2.infn.it}

\author[1]{\fnm{Michele} \sur{Buzzicotti}}\email{michele.buzzicotti@roma2.infn.it}

\author[1]{\fnm{Luca} \sur{Biferale}}\email{luca.biferale@roma2.infn.it}

\author[1]{\fnm{Fabio} \sur{Bonaccorso}}\email{fabio.bonaccorso@roma2.infn.it}

\affil*[1]{\orgdiv{Department of Physics and INFN}, \orgname{University of Rome ``Tor Vergata"}, \orgaddress{\street{Via della Ricerca Scientifica 1}, \city{Rome}, \postcode{00133}, \country{Italy}}}


\abstract{Inference problems for two-dimensional snapshots of rotating turbulent flows are studied.  We perform a systematic quantitative benchmark of point-wise and statistical reconstruction capabilities of the linear Extended Proper Orthogonal Decomposition (EPOD) method, a non-linear Convolutional Neural Network (CNN) and a Generative Adversarial Network (GAN). We attack the important task of inferring one velocity component out of the measurement of a second one, and two cases are studied: (I) both components lay in the plane orthogonal to the rotation axis and (II) one of the two is parallel to the rotation axis. We show that EPOD method works well  only for the former case where both components are strongly correlated, while CNN and GAN always outperform EPOD both concerning point-wise and statistical reconstructions. For case (II), when the input and output data are weakly correlated, all methods fail to reconstruct faithfully the point-wise information.  In this case, only GAN is able to reconstruct the field in a  statistical sense. The analysis is performed using both  standard validation tools based on $L_2$ spatial distance between the prediction and the ground truth and more sophisticated multi-scale analysis using wavelet decomposition. Statistical validation is based on standard Jensen-Shannon divergence between the probability density functions, spectral properties and  multi-scale flatness.}




\maketitle

\section{Introduction}

Understanding and predicting turbulent processes in geophysical and laboratory set-ups is key for many important questions \cite{le1986variational, bell2009godae, krysta2011consistent, storer2022global}. Although observation technologies are progressing, the quality and quantity of available data so far is still inadequate in many respects. The most important limitation are due to measurements sensitivity, spatial and temporal sparsity \cite{shen2015missing, zhang2018missing, militino2019filling}. As a result, different tools have been developed to reconstruct the whole `dense' field from sparse/noisy/limited measurements. Gappy Proper Orthogonal Decomposition (GPOD) is based on a linear reconstruction in eigenmodes \cite{everson1995karhunen}, and it was later improved and applied in \cite{venturi2004gappy} for the simple idealized and  prototypical case of a flow past a circular cylinder with random spatio-temporal gappiness. More recently, inspired by the success of Convolutional Neural Networks (CNNs) in computer vision tasks, a series of proof-of-concepts studies have used a  Generative Adversarial Network (GAN) \cite{goodfellow2014generative}  to reconstruct two-dimensional (2D) snapshots of three-dimensional (3D) rotating turbulence with a large gap \cite{buzzicotti2021reconstruction, li2022data}. In our previous work \cite{li2022data}, POD- and GAN-based reconstructions have been systematically compared in terms of the point-wise error and turbulent statistical properties, for gaps with different sizes and geometries. For super-resolution tasks, CNN and GAN have been applied to recover high-resolution laminar or turbulent flows from low-resolution coarse data in space and time \cite{liu2020deep, subramaniam2020turbulence, fukami2021machine, kim2021unsupervised}. Moreover, CNN or GAN have been recently proposed to reconstruct the 3D velocity fields from several 2D sections \cite{matsuo2021supervised} or a cross-plane of unpaired 2D velocity observations \cite{yousif2022deep}. 

In this paper, we deal with different inference problems, connected to gappiness in the diversity of physical fields that can be measured. In particular, we ask how much one can make use of the measurement of one component of the 3D turbulent velocity field to predict another one. The problem is of course important for many field and laboratory applications where sensors/observations can collect fluctuations only in some preferred directions, e.g. Particle Image Velocimetry (PIV) \cite{adrian2011particle, di2022reconstructing} and wind observation from satellite infrared images \cite{liberzon2004xpiv, elsinga2006tomographic, sheng2008using, dabas2010observing, lin2020ocean}. Inference problems can be attacked by Extended Proper Orthogonal Decomposition (EPOD) using linear decomposition on a predefined basis of independent functions \cite{boree2003extended}. Recently, for a turbulent open-channel flow, it has been shown that the 2D velocity-fluctuation fields at different wall-normal locations can be predicted with a good statistical accuracy from the wall-shear-stress components and the wall pressure with a fully non-linear CNN \cite{guastoni2021convolutional} and GAN \cite{guemes2021coarse}. Here we ask a more complex multi-objective question, checking how much one can infer the unkown field in terms of (i) the  point-to-point $L_2$ error and (ii) the statistical properties, e.g. on the basis of Jensen-Shannon (JS) divergence between the probability density function (PDF) of the ground truth and the one of the predicted velocity component.

\begin{figure}
	\centering
	\includegraphics[width=1.0\linewidth]{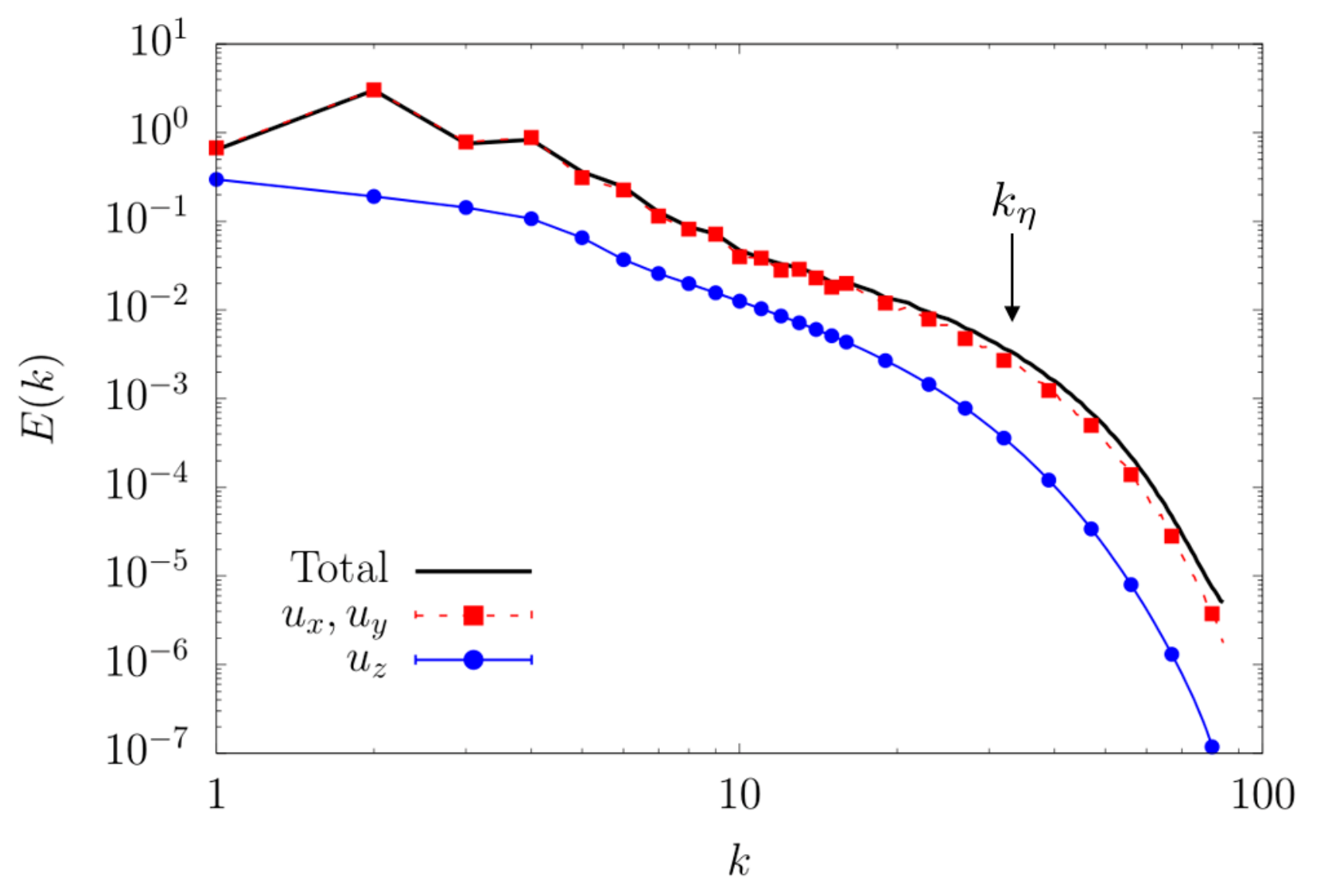}
	\caption{Energy spectra of different velocity components and the total energy spectrum from the TURB-Rot database \cite{biferale2020turb}. The Kolmogorov dissipative wave number, $k_\eta=32$, is determined as the scale where the total energy spectrum starts to decay exponentially. }
	\label{fig:Dataset_spectra}
\end{figure}

\begin{figure}
	\centering
	\includegraphics[width=1.0\linewidth]{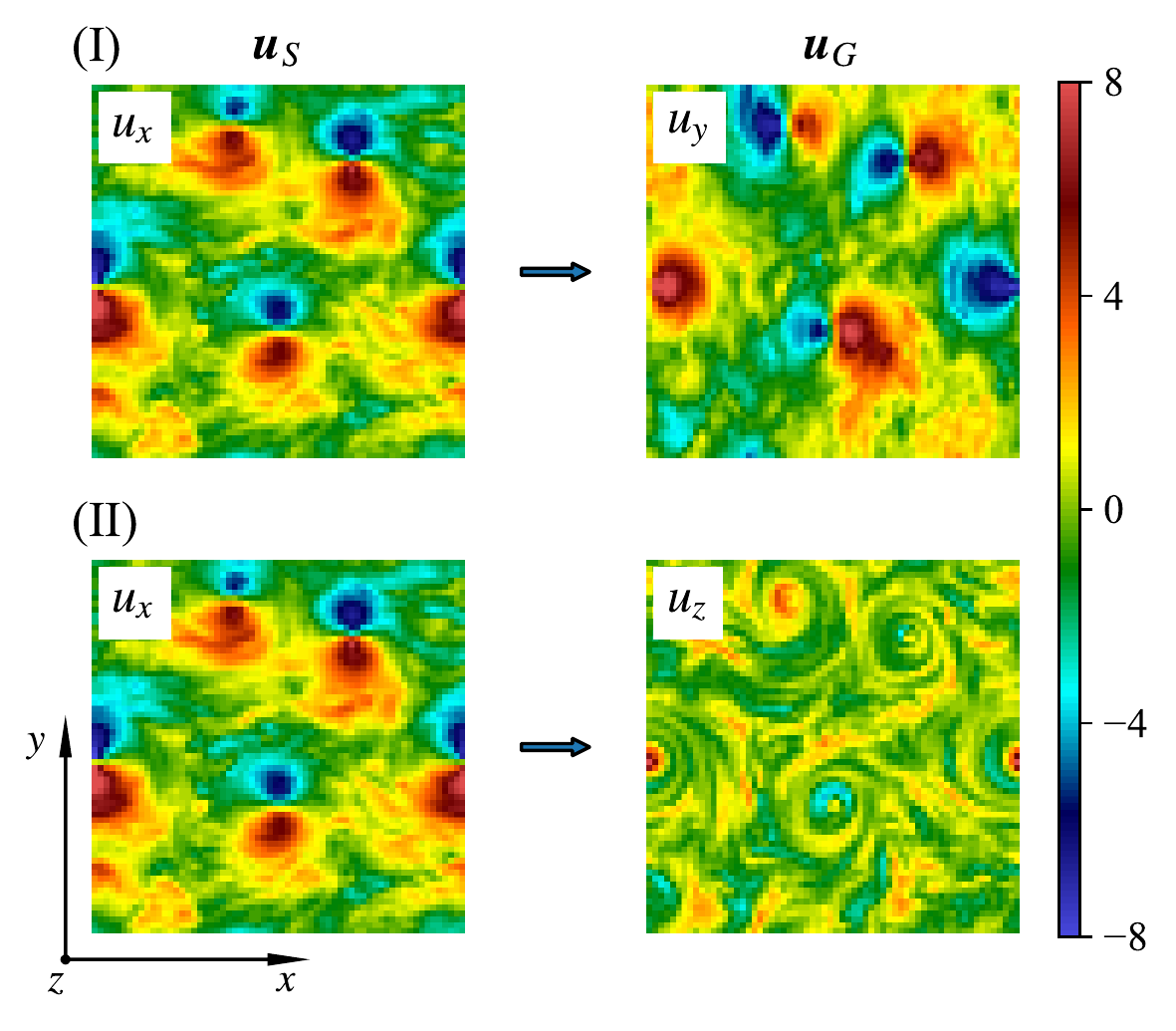}
	\caption{Examples of two inference tasks on 2D slices of 3D turbulent rotating flows: (I) Using $u_x$ to infer $u_y$ and (II) using $u_x$ to infer $u_z$. The rotation axis is along $z$-direction. Note that the in-plane components, $u_x$ and $u_y$, are strongly correlated among each other while the out-plane component, $u_z$, is less correlated with both $u_x$ and $u_y$. Here, $\bm u_S$ and $\bm u_G$ respectively represent the measured quantities and those to be inferred.}
	\label{fig:Inference_task}
\end{figure}

In order to be specific with some important applications, we concentrate on the case of 3D turbulent flows under rotation, a system with plenty of physical interest and of high relevance to geophysical and engineering problems \cite{cho2008atmospheric, le2017inertial, dumitrescu2004rotational}. Under strong rotation, the flow tends to become quasi-2D, with the formation of large-scale coherent vortical structures parallel to the rotation axis \cite{chen2005resonant, davidson2015turbulence, buzzicotti2018energy}. Moreover, strong non-Gaussian, 3D and intermittent fluctuations exist at small scales because of the forward energy cascade \cite{mininni2009scale, godeferd2015structure, alexakis2018cascades, di2020phase}. The tendency towards quasi-2D configurations implies that the out-plane velocity component, $u_z$, behaves close to a passive scalar advected by the in-plane components, $u_x$ and $u_y$. This can be further shown in Fig. \ref{fig:Dataset_spectra} with the energy spectra of different velocity components from the database used in this study. It is obvious that the energy of $u_x$ or $u_y$ is dominant compared with that of $u_z$ over the whole range of scales, which indicates that $u_z$ is almost decoupled from the leading dynamics. Therefore, $u_x$ and $u_y$ are well correlated with each other while $u_z$ is expected to show less correlations  (Fig. \ref{fig:Inference_task}). In this work, we assess the potential of EPOD, CNN and GAN methods to infer the velocity components on 2D slices perpendicular to the rotation axis. We limit the analysis to the scenario where only instantaneous measurements are provided without any temporal sequence information. Two tasks are studied with different difficulties as shown in Fig. \ref{fig:Inference_task}: (I) Using one in-plane component, $u_x$, to predict the other, $u_y$, and (II) using an in-plane component, $u_x$, to predict the out-plane one, $u_z$.

The organization of this paper is as follows. In Sect. \ref{sec:Methodology}, we describe the databset used to evaluate the inference tools and introduce the EPOD and GAN-based methods. The CNN used in this study is taken from the generator part of the GAN. Results of above tools on the two inference tasks are given in Sect. \ref{sec:Results}. Finally, we present the conclusions in Sect. \ref{sec:Conclusion}.

\section{Methodology}\label{sec:Methodology}

\subsection{Dataset}

To evaluate different tools on the inference of velocity components, we use the TURB-Rot database \cite{biferale2020turb}, obtained from Direct Numerical Simulations (DNS) of the Navier-Stokes equations for incompressible rotating fluid in a 3D periodic domain, which can be written as
\begin{equation}\label{equ:mmt}
	\frac{\pt\bm{u}}{\pt t}+\bm{u}\cdot\bnabla\bm{u}+2\bm{\mathit{\Omega}}\times\bm{u}=-\frac{1}{\rho}\bnabla\tilde{p}+\nu\Delta\bm{u}+\bm{f},
\end{equation}
where $\bm{u}$ is the incompressible velocity, $\bm{\mathit{\Omega}}=\mathit{\Omega}\hat{\bm{z}}$ is the rotation vector, $\tilde{p}$ is the modified pressure in the rotating frame,the forcing $\bm{f}$ acts at scales around $k_f=4$ and it is generated from a second-order Ornstein-Uhlenbeck process \cite{sawford1991reynolds, buzzicotti2016lagrangian}.  To enlarge the inertial range, the dissipation $\nu\Delta\bm{u}$ is modeled by a hyperviscous term $\nu_h\nabla^4\bm{u}$. Moreover, a linear friction term $\beta\Delta^{-1}\bm{u}$ is used at large scales to reduce the formation of a large-scale condensate \cite{alexakis2018cascades}. The Rossby number in the stationary regime is $Ro={\cal E}^{1/2}/(\mathit{\Omega}/k_f)\approx0.1$, where ${\cal E}$ is the flow kinetic energy. The Kolmogorov dissipative wave number, $k_\eta=32$, is chosen as the scale where the total energy spectrum $E(k)$ starts to decay exponentially (Fig. \ref{fig:Dataset_spectra}). Given the domain length $L_0$, the integral length scale is $L={\cal E}/\int kE(k)\,\mathrm{d}k\approx0.15L_0$ and the integral time scale is $T=L/{\cal E}^{1/2}\approx0.185$. Refer to \cite{biferale2020turb} for more details of the simulation.

Although all inference tools in this study can be applied to 3D data, here we restrict to 2D horizontal slices in order to make contact with the geophysical observation, PIV experiments and to restrict the amount of data to be used for training (see \cite{subramaniam2020turbulence, fukami2021machine, matsuo2021supervised, yousif2022deep} for a few applications to 3D datasets). To extract data from the simulation, we first sampled snapshots of the full 3D velocity field during the stationary regime. Snapshots are chosen with a large temporal interval $\Delta t_s=5.41T$ to decrease their correlations in time. There are 600 snapshots sampled over $3243T$ at early times for training and validation, while 160 snapshots sampled over $865T$ at later times. The time separation between the two samplings for training/validation and testing is more than $3459T$. Second, the resolution of the sampled fields is downsized from $256^3$ to $64^3$ by a Galerkin truncation in Fourier space:
\begin{equation}
	\bm{u}(\bm{x})=\sum_{\lVert\bm{k}\rVert\le k_\eta}\hat{\bm{u}}(\bm{k})\mathrm{e}^{\mathrm{i}\bm{k}\cdot\bm{x}},
\end{equation}
where the cut-off wave number is chosen to be the Kolmogorov dissipative wave number, such as to eliminate only fully dissipative degrees of freedom where the flow can be approximated to be linear with a good approximation. This is needed to balance the request to reduce the amount of data to be analyzed, without reducing the complexity of the task. For each downsized configuration, we selected 16 $x$-$y$ planes at different $z$ levels and each plane is augmented to 11 (for training/validation) or 8 (for testing) different ones using the periodic boundary condition \cite{biferale2020turb}. After independent random shuffles of the two sets of planes, the dataset is in Train/Validation/Test split as follows, 84480/10560/20480, corresponding to 73.1\%, 9.1\% and 17.7\% respectively.

\subsection{EPOD inference}

Denote vector $\bm{u}_S$ as the measured quantities and $\bm{u}_G$ as the quantities to be inferred, which are defined on the spatial 2D domain, $\Omega$. Note that $\bm{u}_S$ or $\bm{u}_G$ becomes a scalar when there is only one quantity measured or to be inferred. For the inference task (I), one uses $u_x$ to predict $u_y$, which can be expressed as
\begin{equation}
	\bm{u}_S\colon u_x\to\bm{u}_G\colon u_y,
\end{equation}
and for the inference task (II),
\begin{equation}
	\bm{u}_S\colon u_x\to\bm{u}_G\colon u_z,
\end{equation}
see Fig. \ref{fig:Inference_task}.
%\blue{In the rest of All the inference methodologies are capable of the general case, thus are introduced with $\bm{u}_S$ and $\bm{u}_G$ in the following text.}
The first step of EPOD method is to compute the correlation matrix
\begin{equation}
\utilde{\mathcal{R}}_S(\bm{x},\bm{y})=\langle\bm{u}_S(\bm{x})\bm{u}_S(\bm{y})^T\rangle,
\end{equation}
where with $\langle\cdot\rangle $ we denote an average over the configurations  selected for the  training set. Then we solve the eigenvalue problem
\begin{equation}
\int_\Omega\utilde{\mathcal{R}}_S(\bm{x},\bm{y})\cdot\bm{\phi}_S^{(n)}(\bm{y})\,\mathrm{d}\bm{y}=\sigma_n\bm{\phi}_S^{(n)}(\bm{x}),
\end{equation}
where $n=1,\ldots,N_\Omega$ and $N_\Omega$ is the number of physical points in the domain $\Omega$, to obtain the eigenvalues $\sigma_n$ and the POD eigenmodes $\bm{\phi}_S^{(n)}(\bm{x})$. It is easy to realize that any realization of the observed fields can be decomposed as
\begin{equation}
	\bm{u}_S(\bm{x})=\sum_{n=1}^{N_\Omega}b^{(n)}_S\bm{\phi}_S^{(n)}(\bm{x}),
\end{equation}
where the POD coefficients are given by the inner product:
\begin{equation}\label{equ:coe}
	b_S^{(n)}=\int_\Omega\bm{u}_S(\bm{x})\cdot\bm{\phi}_S^{(n)}(\bm{x})\,\mathrm{d}\bm{x}.
\end{equation}
Furthermore, exploiting the orthogonality of the POD eigenmodes one can also write the following identity \cite{boree2003extended}:
\begin{equation}\label{equ:POD}
	\bm{\phi}_S^{(n)}(\bm{x})=\langle b^{(n)}_S\bm{u}_S(\bm{x})\rangle/\sigma_n.
\end{equation}
The idea of the {\it Extended} POD modes is to generalize the exact relation (\ref{equ:POD}) also to the case where on the r.h.s. we use the quantities to be inferred, i.e. by replacing $\bm{u}_S$ with $\bm{u}_G$ but keeping using the same coefficients $b_n$ defined in (\ref{equ:coe}):
\begin{equation}\label{eq:ext}
	\bm{\phi}_E^{(n)}(\bm{x})=\langle b_S^{(n)}\bm{u}_G(\bm{x})\rangle/\sigma_n.
\end{equation}
 Once completed the {\it training} protocol and defined the set of EPOD modes (\ref{eq:ext}) one can start the inferring procedure for any new configuration outside the training set, by taking the measured components $\bm{u}_S(\bm{x})$, calculating the coefficients (\ref{equ:coe}) and defining the predicted/inferred field as
\begin{equation}\label{eq:pre_epod}
	\bm{u}_G^{(p)}(\bm{x})=\sum_{n=1}^{N_\Omega}b_S^{(n)}\bm{\phi}_E^{(n)}(\bm{x}).
\end{equation}
$\bm{u}_G^{(p)}$  represents the  prediction given by an inference method (here EPOD). Note that $b_S^{(n)}$ in (\ref{eq:pre_epod}) is calculated out of (\ref{equ:POD}) with measurements of a testing data, which is different from $b_S^{(n)}$ in (\ref{eq:ext}) for the training set.

\subsection{GAN-based inference with context encoders}

\begin{figure}
	\centering
	\includegraphics[width=1.0\linewidth]{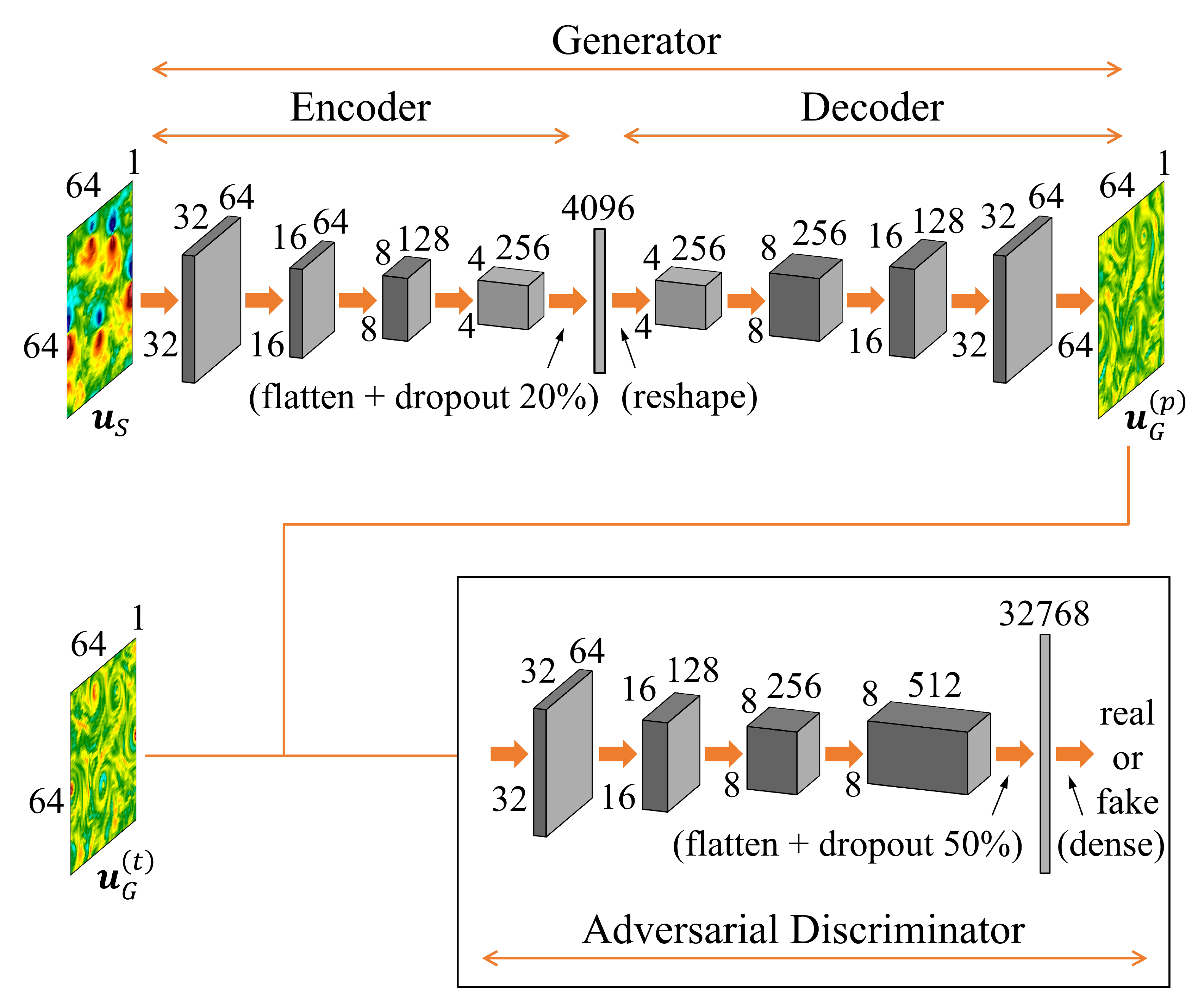}
	\caption{Architecture of generator and discriminator for the velocity inference tasks. Each convolution (up-convolution) layer has a kernel size of 4 followed by a Leaky Rectified Linear Unit (ReLU) activation function, except that the last layer of the generator uses a Tanh activation function. We can increase channels in the layer $\bm{u}_S$ or the layers $\bm{u}_G^{(p)}$ and $\bm{u}_G^{(t)}$ when there are more than one components/quantities measured or to be inferred.}
	\label{fig:GAN_schematic_diagram}
\end{figure}

GAN \cite{goodfellow2014generative} has been proposed to generate 2D images out of random input sets. The architecture consists of two neural networks: a generator and a discriminator. The generator tries to create new data that is -statistically- similar to the training data, while the discriminator tries to distinguish between the real and the generated data. Both networks are trained together in a game-like fashion where the generator tries to fool the discriminator, and the discriminator tries to accurately distinguish between real and generated data. Over the learning phase, the generator gets better at creating realistic data that can fool the discriminator, leading to high-quality generated data. 
For the task here attacked, inferring  data out of some input configuration, the typical GAN architecture must be slightly adapted, using a context encoder \cite{pathak2016context} which  takes as input the measured data (instead of a random vector) and adds a second loss measuring the $L_2$ point-to-point distance between the GAN output and the ground truth configuration.   These architectures have already been used  to reconstruct gappy 2D turbulent configurations in  our previous works \cite{buzzicotti2021reconstruction, li2022data}. For the inference tasks of velocity components here studied, the GAN architecture is shown in Fig. \ref{fig:GAN_schematic_diagram}.
The generator {is a functional $GEN(\cdot)$} first encoding the supplied measurements, $\bm{u}_S$, to a latent feature representation, and second it generalizes by a decoder the latent representation to produce a candidate for the missing measurements, ${GEN(\bm{u}_S)=}\bm{u}_G^{(p)}$, while $\bm{u}_G^{(t)}$ is the missing true turbulent configuration. {The discriminator plays as a `referee' functional $D(\cdot)$ which takes either $\bm{u}_G^{(t)}$ or $\bm{u}_G^{(p)}$ and outputs the probability that the supplied configuration is sampled from the true turbulent dataset.} The loss function of the generator is
\begin{equation}\label{eq:GEN}
	\mathcal{L}_{GEN}=(1-\lambda_{adv})\mathcal{L}_\mathrm{MSE}+\lambda_{adv}\mathcal{L}_{adv},
\end{equation}
where the $L_2$ loss
\begin{equation}\label{eq:L_MSE}
	\mathcal{L}_\mathrm{MSE}=\langle \frac{1}{A_\Omega} \int_I  \|\bm{u}_G^{(p)}(\bm{x})-\bm{u}_G^{(t)}(\bm{x})\|^2 \,\mathrm{d}\bm{x} \rangle
\end{equation}
is defined as the mean squared error (MSE) averaged over the spatial domain of area $A_\Omega$. The hyper-parameter $\lambda_{adv}$ is called the adversarial ratio and {the adversarial loss is 
\begin{align}
    \label{eq:L_adv}
	\mathcal{L}_{adv}&=\langle\log(1-D(\bm{u}_G^{(p)}))\rangle \nonumber \\
    &=\int p(\bm{u}_S)\log[1-D(GEN(\bm{u}_S))]\,\mathrm{d}\bm{u}_S \nonumber \\
    &=\int p_p(\bm{u}_G)\log(1-D(\bm{u}_G))\,\mathrm{d}\bm{u}_G,
 \end{align}
where $p(\bm{u}_S)$ is the probability distribution of the known input measurements in the training set and $p_p(\bm{u}_G)$ is the probability distribution of the predicted fields from the generator. The discriminator is trained simultaneously with the generator to maximize the cross entropy between the ground truth and generated samples
\begin{align}
    \label{equ:L_DIS}	
    &\mathcal{L}_{DIS}=\langle\log(D(\bm{u}_G^{(t)}))\rangle+\langle\log(1-D(\bm{u}_G^{(p)}))\rangle \nonumber \\
    &=\int[p_t(\bm{u}_G)\log(D(\bm{u}_G))+ \nonumber \\
    &\qquad\qquad\qquad p_p(\bm{u}_G)\log(1-D(\bm{u}_G))]\,\mathrm{d}\bm{u}_G.
\end{align}
Here $p_t(\bm{u}_G)$ is the probability distribution of the real -ground truth- fields, $\bm{u}_G^{(t)}$. It is possibile to show that the generative-adversarial training with $\lambda_{adv}=1$ in (\ref{eq:GEN}) minimizes the JS divergence between the generated and the true probability distributions, $\mathrm{JSD}(p_t\parallel p_p)$ \cite{goodfellow2014generative, nowozin2016f}. } Therefore, the adversarial loss helps the generator to produce predictions with correct turbulent statistics, while the generator, if trained alone ($\lambda_{adv}=0$), would minimize only the $L_2$ loss (\ref{eq:L_MSE}) which is mainly sensitive to the large energy-containing scales. Compared with the linear EPOD, GAN not only takes advantage of the non-linear expression of the generator, but also can optimize a loss considering both the MSE error and the generated probability distribution.

More details of the GAN are provided in Appendix \ref{secA1}, including the architecture, hyper-parameters and the training schedule.

\section{Results}\label{sec:Results}

In this section, we systematically compare EPOD, CNN and GAN-based methods for the two tasks of velocity component inference. The CNN has the same architecture as the generator of the GAN, thus it corresponds to our GAN with $\lambda_{adv}=0$ (without the discriminator); In this way we are also able to judge the importance of adding/removing the objective to have a small JS divergence with the ground truth for our inferred field. For the decision of the adversarial ratio for the GAN, readers can refer to Appendix \ref{secA1}.

To quantify the inference error, we define the normalized mean squared error (MSE) as
\begin{equation}
\label{eq:defMSE}
	\mathrm{MSE}(\bm{u}_G)=\langle\Delta_{\bm{u}_G}\rangle/E_{\bm{u}_G},
\end{equation}
where
\begin{equation}
	\Delta_{\bm{u}_G}=\frac{1}{A(I)}\int_I\|\bm{u}_G^{(p)}(\bm{x})-\bm{u}_G^{(t)}(\bm{x})\|^2\,\mathrm{d}\bm{x}
\end{equation}
is the spatially averaged $L_2$ error for one flow configuration and $\langle\cdot\rangle$
represents now the average over the test data. The normalization factor is defined as
\begin{equation}\label{equ:nor}
E_{\bm{u}_G}=\sigma_G^{(p)}\sigma_G^{(t)},
\end{equation}
where
\begin{equation}\label{equ:sig}
	\sigma_G^{(t)}=\langle\frac{1}{A(I)}\int_I\|\bm{u}_G^{(t)}(\bm{x})\|^2\,\mathrm{d}\bm{x}\rangle^{1/2}
\end{equation}
and $\sigma_G^{(p)}$ is defined similarly. The form of $E_{\bm{u}_G}$ makes sure that a prediction with too small or too large energy gives a large MSE.

We use JS divergence to evaluate the PDF of the inferred velocity components. For two probability distributions, $P(x)$ and $Q(x)$, JS divergence measures their similarity and is defined as
\begin{equation}
	\mathrm{JSD}(P\parallel Q)=\frac{1}{2}\mathrm{KL}(P\parallel M)+\frac{1}{2}\mathrm{KL}(Q\parallel M),
\end{equation}
where $M=\frac{1}{2}(P+Q)$ and
\begin{equation}
 \mathrm{KL}(P\parallel Q)=\int_{-\infty}^\infty P(x)\log\left(\frac{P(x)}{Q(x)}\right)\,\mathrm{d}x
\end{equation}
is the Kullback–Leibler (KL) divergence. If the two probability distributions are close, it gives a small JS divergence, and vice versa.

\subsection{Inference task (I)}
\subsubsection{Prediction at large scales and small scales}
\begin{figure*}
	\centering
	\includegraphics[width=1.0\linewidth]{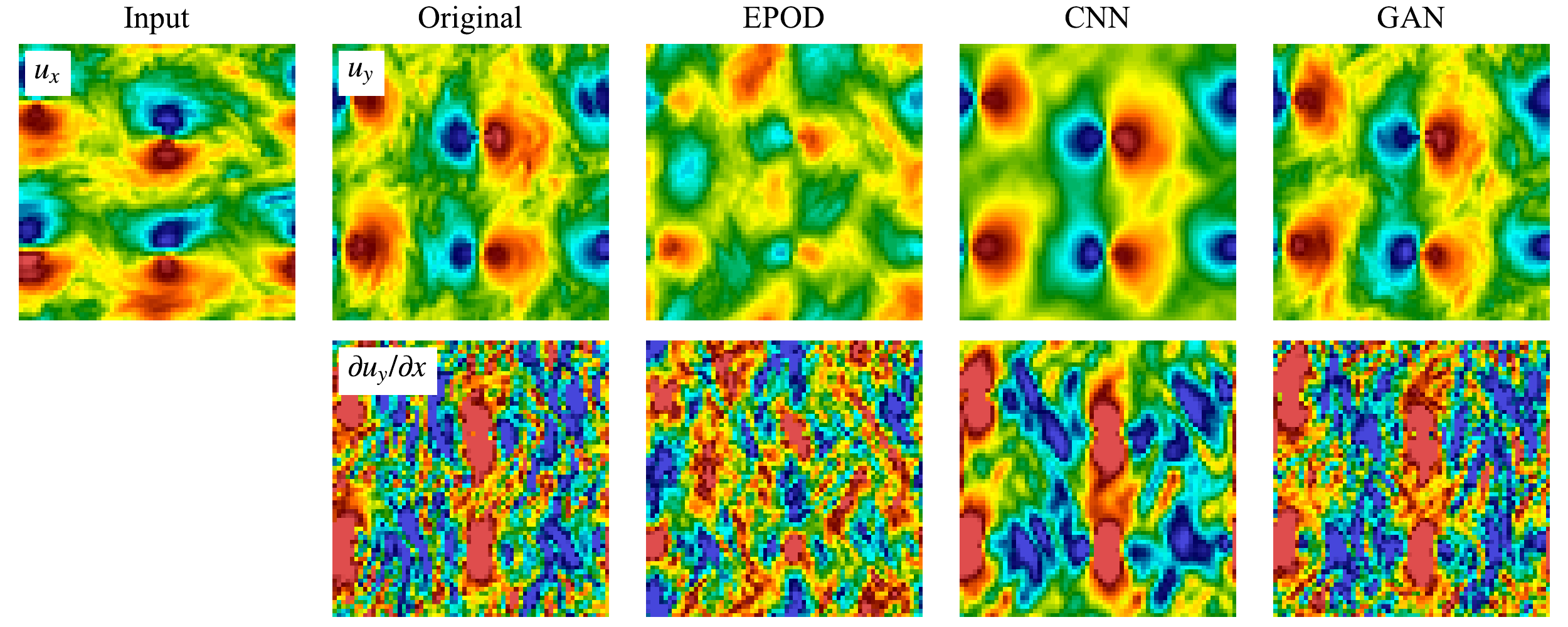}
	\caption{Prediction visualization of the inference task (I) by the different tools for an instantaneous field. In the 1st row, the input velocity component, $u_x$, is shown in the 1st column, while the 2nd to 5th columns show the ground truth and the inferred velocity component, $u_y$, obtained from EPOD, CNN and GAN. The corresponding gradient in $x$-direction, $\pt u_y/\pt x$, is shown in the 2nd row.}
	\label{fig:Reconstruction-uy-ux}
\end{figure*}

For the inference task (I), in Fig. \ref{fig:Reconstruction-uy-ux} we present an inference experiment using a test data never showed during training. The aim is to use the velocity component $u_x$ (1st row, 1st column) to predict the velocity component $u_y$, of which the ground truth and predictions from the different tools are shown in the other columns of the 1st row. In the 2nd row we show their gradients in $x$-direction, $\pt u_y/\pt x$. EPOD only keeps the correlated part with the given information and the prediction looks meaningful because of the correlation between $u_x$ and $u_y$. With the capability of expressing non-linear correlations, CNN predicts better results which are close to the ground truth. However,  CNN predictions are blurry without the high-frequency information, while GAN can generate realistic turbulent configurations with the benefit of adversarial training.

\begin{figure}
	\centering
	\includegraphics[width=1.0\linewidth]{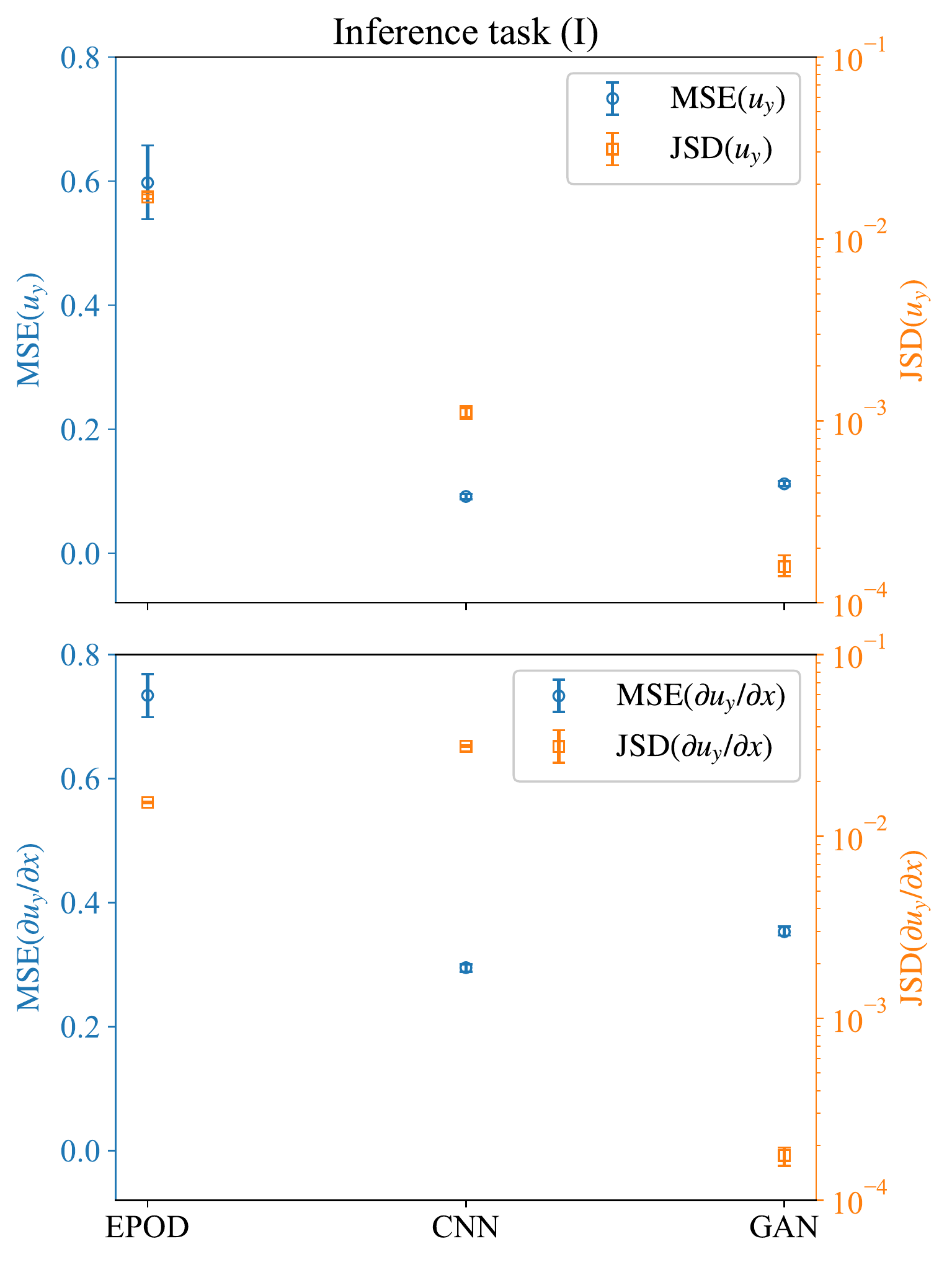}
	\caption{MSE (left y-axis) and JS divergence (right y-axis) between PDFs for the velocity component to be inferred, $u_y$, and its gradient, $\pt u_y/\pt x$. Results are obtained from EPOD, CNN and GAN for the inference task (I).}
	\label{fig:MSE-uy-ux}
\end{figure}

\begin{figure}
	\centering
	\includegraphics[width=1.0\linewidth]{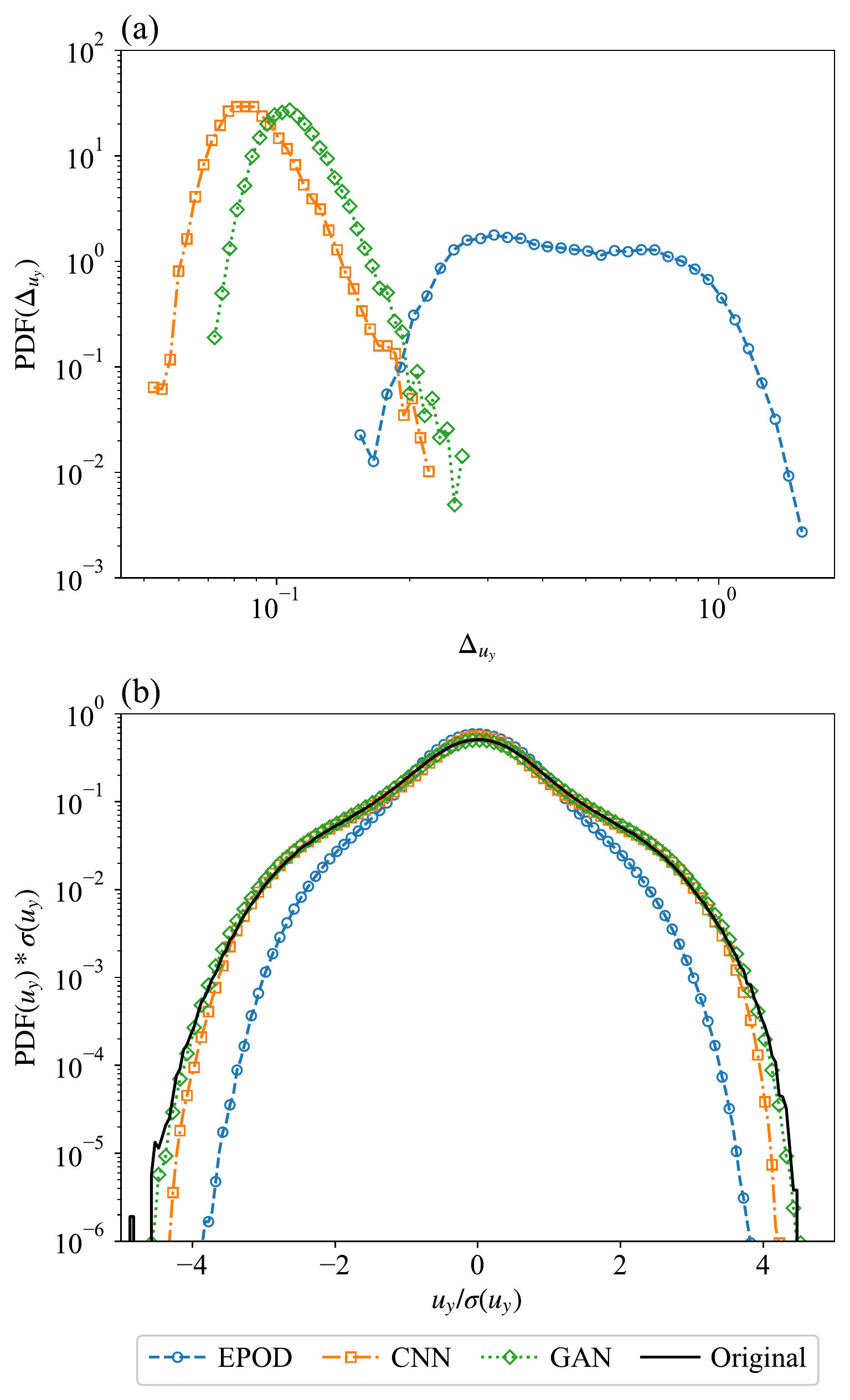}
	\caption{(a) PDFs of the spatially averaged $L_2$ error over different configurations and (b) PDFs of the predicted and the original velocity components, $u_y$, where $\sigma(u_y)$ is the standard deviation of the original data. Results are obtained from EPOD, CNN and GAN for the inference task (I).}
	\label{fig:PDF_L2_c-PDF-uy-ux}
\end{figure}

To analyze the predictions quantitatively, Fig. \ref{fig:MSE-uy-ux}(top) shows the $\mathrm{MSE}(u_y)$ and the JS divergence between PDFs of the original and the predicted $u_y$, which is denoted as $\mathrm{JSD}(u_y)=\mathrm{JSD}(\mathrm{PDF}(u_y^{(t)})\parallel\mathrm{PDF}(u_y^{(p)}))$. We divide the test data into batches of size 128 (or 2048), calculate the MSE (or JS divergence) over these batches and we indicate with the error bound its range of fluctuation. The EPOD approach has a $\mathrm{MSE}(u_y)$ around 0.6, while CNN and GAN have smaller values around 0.1.
%\begin{table*}[h]
%	\begin{center}
%		\begin{minipage}{\textwidth}
%			\caption{\red{can you replace the table with a plot? and error bars?} MSE and JS divergence between PDFs for the velocity component to be inferred, $u_y$. Results are obtained from the different tools for the inference task (i).}\label{tab:task1}%
%			\begin{tabular}{@{}cccccc@{}}
%				\toprule
%				& GPOD (DR) & GPOD (Lasso) & EPOD & CNN & GAN \\
%				\midrule
%				$\mathrm{MSE}(u_y)$ & $5.444_{-0.242}^{+0.190}$ & $2.290_{-0.123}^{+0.132}$ & $0.5910_{-0.0581}^{+0.0596}$ & $0.09048_{-0.00445}^{+0.00389}$ & $0.1106_{-0.0037}^{+0.0047}$ \\
%				$\mathrm{JSD}(u_y)$ & $3.89_{-0.04}^{+0.06}\times10^{-1}$ & $6.79_{-1.07}^{+1.25}\times10^{-2}$ & $1.70_{-0.04}^{+0.06}\times10^{-2}$ & $1.12_{-0.10}^{+0.05}\times10^{-3}$ & $1.58_{-0.18}^{+0.24}\times10^{-4}$ \\
%				\botrule
%			\end{tabular}
%		\end{minipage}
%	\end{center}
%\end{table*}
Fig. \ref{fig:PDF_L2_c-PDF-uy-ux} shows the PDF of the spatially averaged $L_2$ error, $\Delta_{u_y}$, over different configurations, the peaks of which give consistent results with MSEs. The JS divergence of the inferred component can be explained together with Fig. \ref{fig:PDF_L2_c-PDF-uy-ux}(b), which shows PDFs of the predicted velocity component compared with the original data. EPOD has the largest value of $\mathrm{JSD}(u_y)$ and the predicted $\mathrm{PDF}(u_y)$ has the correct shape but deviates from the original one at around $\lvert u_y\rvert=\sigma(u_y)$, where $\sigma(u_y)$ represents the standard deviation of the original data. CNN and GAN generate nearly perfect PDFs with small values of $\mathrm{JSD}(u_y)$. Moreover, GAN is better with the help of the discriminator, which also slightly increases $\mathrm{MSE}(u_y)$.
Fig. \ref{fig:MSE-uy-ux}(bottom) shows the MSE and JS divergence between PDFs for the gradient of $u_y$, $\pt u_y/\pt x$. The $\mathrm{MSE}(\pt u_y/\pt x)$ behaves similarly as $\mathrm{MSE}(u_y)$ that EPOD gives a large value while  CNN and GAN have smaller close values. However, for the gradient both EPOD and CNN give large values of JS divergence and only GAN predicts a small value, indicating the importance of adversarial training on the high-order quantities.

\subsubsection{Multi-scale prediction error}
\begin{figure}
    \centering
    \includegraphics[width=0.9\linewidth]{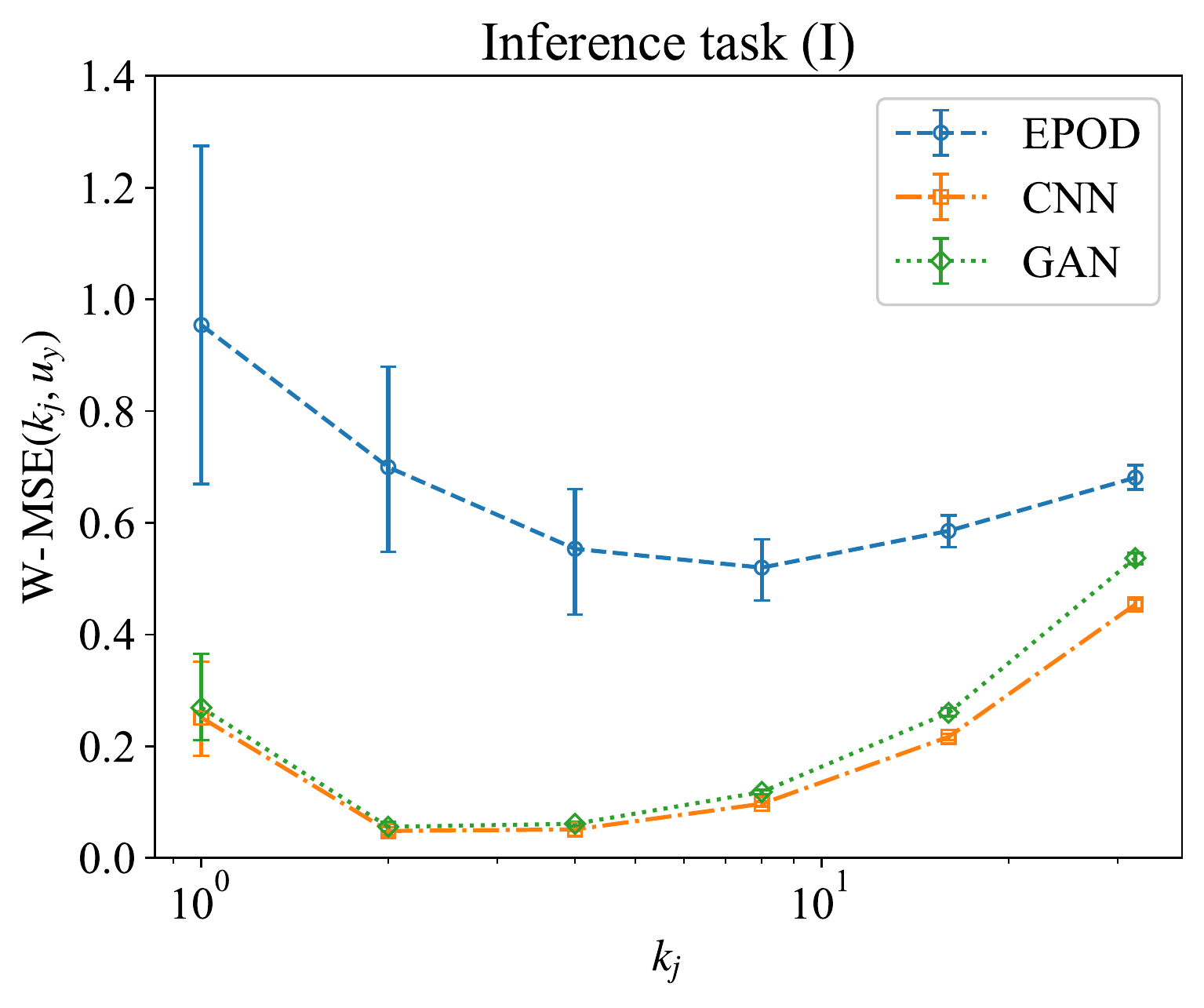}
    \caption{W-MSE for the velocity component to be inferred, $u_y$, at different wave numbers $k_j$. Results are obtained from the different tools for the inference task (I).}
    \label{fig:W_MSE-uy-ux}
\end{figure}

Here,  we evaluate the prediction from a multi-scale perspective with the help of wavelet analysis \cite{farge1992wavelet, benzi1993random, bettega2009wavelet}. For a field defined on a uniform grid of size $2^N\times2^N$, 2D wavelet decomposition gives that
\begin{equation}
    u_y(\bm{x})=\bar{u}_y+\sum_{j=0}^{N-1}u_y^{(k_j)}(\bm{x}),
\end{equation}
where $\bar{u}_y$ is the mean value and
\begin{equation}
    u_y^{(k_j)}(\bm{x})=\sum_{i_x=0}^{2^j-1}\sum_{i_y=0}^{2^j-1}\sum_{\sigma}c_{j,i_x,i_y}^{(\sigma)}\psi_{j,i_x,i_y}^{(\sigma)}(\bm{x})
\end{equation}
is the wavelet contribution at wave number $k_j=2^j$, corresponding to the length scale $1/k_j$. Given that $\sigma\in\lbrace x,y,d\rbrace$, $c_{j,i_x,i_y}^{(\sigma)}$ is the wavelet coefficient and
\begin{equation}
    \begin{aligned}
        \psi_{j,k_x,k_y}^{(x)}(x,y)=\psi_{j,k_x}(x)\phi_{j,k_y}(y), \\
        \psi_{j,k_x,k_y}^{(y)}(x,y)=\phi_{j,k_x}(x)\psi_{j,k_y}(y), \\
        \psi_{j,k_x,k_y}^{(d)}(x,y)=\psi_{j,k_x}(x)\psi_{j,k_y}(y),
    \end{aligned}
\end{equation}
where $\phi(\cdot)$ and $\psi(\cdot)$ are the Haar scaling function and associated wavelet, respectively. To measure the inference error at different scales, we define the normalized wavelet mean squared error (W-MSE) as
\begin{equation}
    \text{W-MSE}(k_j,u_y)=\mathrm{MSE}(u_y^{(k_j)}),
\end{equation}
which is the MSE between wavelet contributions at $k_j$ of the  predicted and original data, following the definition (\ref{eq:defMSE}). We stress that the normalization factor given by (\ref{equ:nor}) and (\ref{equ:sig}) is different for different $k_j$. Fig. \ref{fig:W_MSE-uy-ux} displays the W-MSE obtained from different methods. It shows that at all scales EPOD has the largest prediction error. CNN and GAN produce close values of W-MSE, which is large at small scales where $k_j=2^6$, while varies between 0 to 0.4 at $1\le k_j\le 2^5$. The wave number $k_j$ with the minimum W-MSE, i.e. the scale where the maximum percentage prediction is reached, corresponds to the scale in correspondence of the maximum of the spectrum, as shown by comparing with Fig. \ref{fig:Spectrum-Flatness-uy-ux}(a).

\subsubsection{Spectra and Flatness}
\begin{figure}
	\centering
	\includegraphics[width=1.0\linewidth]{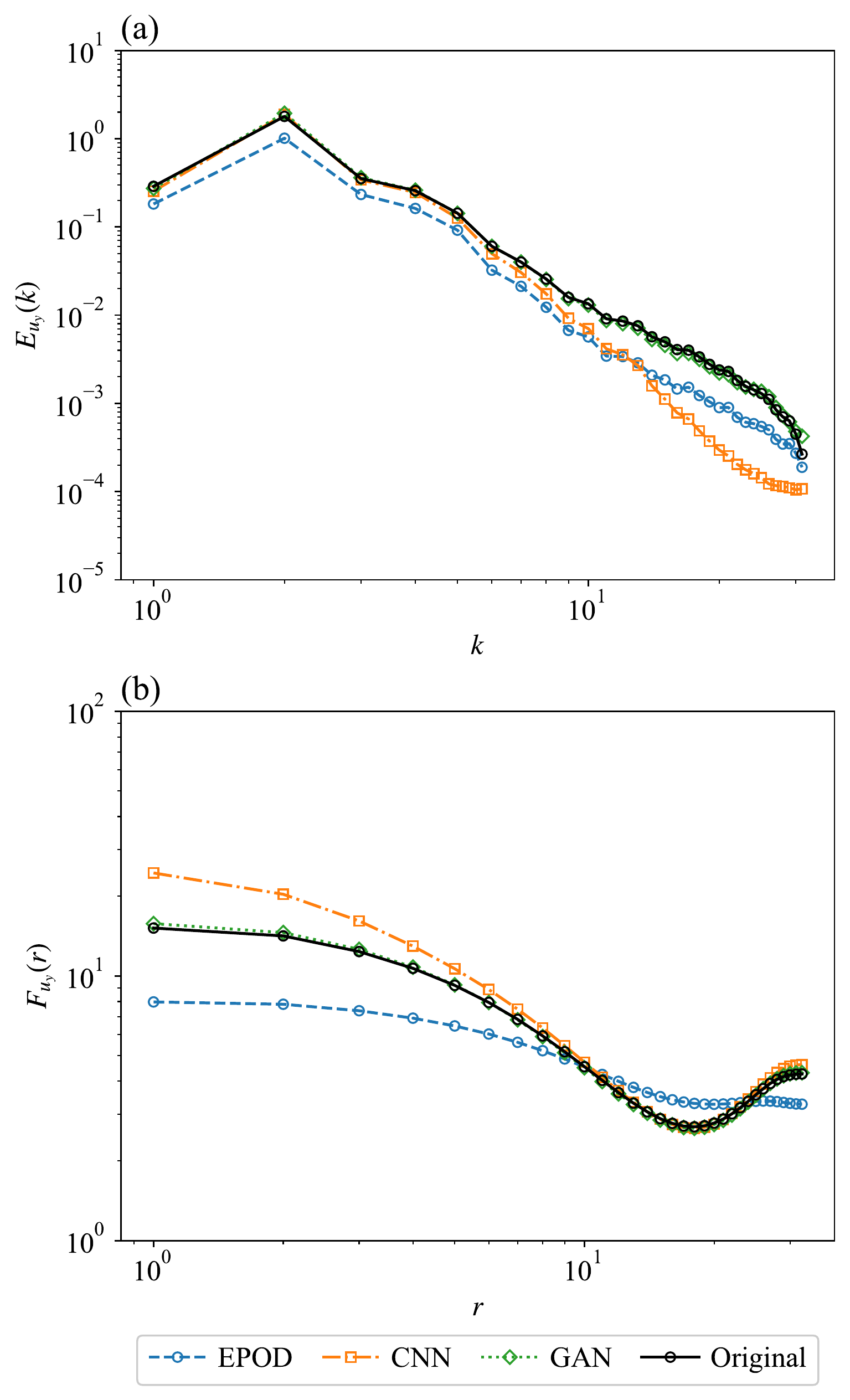}
	\caption{(a) The energy spectrum and (b) the flatness of predictions and the ground truth of the velocity component, $u_y$. The unit of $r$ is the grid width, $w_g=2\pi/64$. Results are obtained from the different tools for the inference task (I).}
	\label{fig:Spectrum-Flatness-uy-ux}
\end{figure}

To further study the statistical properties of different predictions, in Fig. \ref{fig:Spectrum-Flatness-uy-ux}(a) we have measured the energy spectrum for the different inferred fields, namely
\begin{equation}
	E_{u_y}(k)=\sum_{k\le\|\bm{k}\|<k+1}\frac{1}{2}\langle\hat{u}_y(\bm{k})\hat{u}_y^\ast(\bm{k})\rangle,
\end{equation}
where $\bm{k}=(k_x, k_y)$ is the wave number, $\hat{u}_y(\bm{k})$ is the Fourier transform of $u_y(\bm{x})$ and $\hat{u}_y^\ast(\bm{k})$ is its complex conjugate. Since EPOD only extracts the correlated part with the supplied information, it predicts an energy spectrum with a similar shape but with smaller energy at all wave numbers compared with the original one. CNN benefits from the non-linear properties and predicts the correct energy spectrum at small wave numbers. With the help of the discriminator, GAN can predict close energy spectrum to the original one at all wave numbers.
Fig. \ref{fig:Spectrum-Flatness-uy-ux}(b) plots the flatness
\begin{equation}
	F_{u_y}(r)=\langle(\delta_r u_y)^4\rangle/\langle(\delta_r u_y)^2\rangle,
\end{equation}
where $\delta_r u_y=u_y(x+r,y)-u_y(x,y)$. The results are consistent with those of energy spectrum where EPOD fails at all scales, CNN predicts correct flatness at large scales and GAN has satisfying results at all scales.

\subsection{Inference task (II)}
\begin{figure*}
	\centering
	\includegraphics[width=1.0\linewidth]{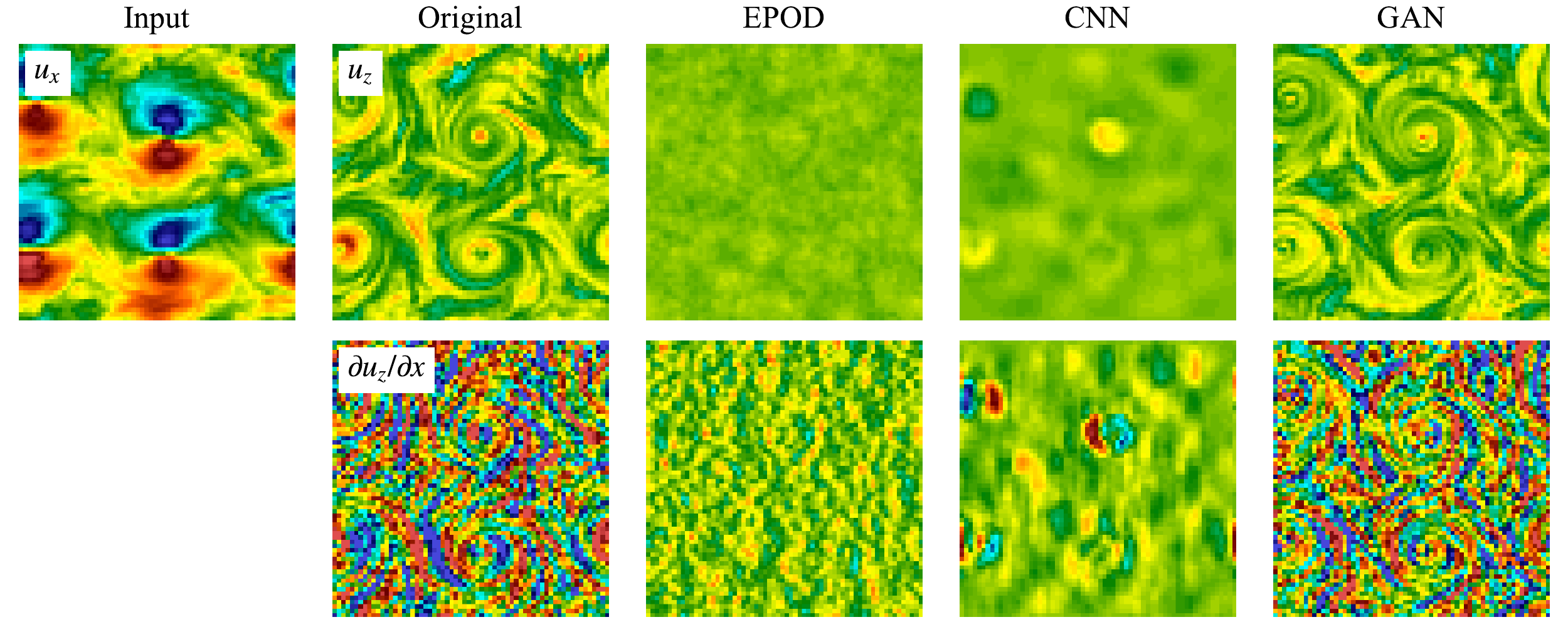}
	\caption{Prediction visualization of the inference task (II) by the different tools for an instantaneous field. In the 1st row, the input velocity component, $u_x$, is shown in the 1st column, while the 2nd to 5th columns show the ground truth and the inferred velocity component, $u_z$, obtained from EPOD, CNN and GAN. The corresponding gradient in $x$-direction, $\pt u_z/\pt x$, is shown in the 2nd row.}
	\label{fig:Reconstruction-uz-ux}
\end{figure*}

Now we move to the inference task (II) which is more challenging as illustrated in the introduction. Fig. \ref{fig:Reconstruction-uz-ux} presents an inference experiment similar as Fig. \ref{fig:Reconstruction-uy-ux} but for the task (II). It is obvious that GAN predictions are realistic and well correlated with the ground truth. However, although the predicted structures are correct, GAN predictions can have wrong values which can be even with opposite signs in some vortices (5th column). The EPOD method is not able to infer meaningful results due to the complexity of the task. The 4th column shows that CNN predicts some blurry blobs that can capture the positions of the large-scale vortices, but cannot correctly predict their signs. This limited capability of CNN should be the reason why GAN cannot predict correct signs of vortices either, as GAN uses the same CNN as its generator.

\begin{figure}
	\centering
	\includegraphics[width=1.0\linewidth]{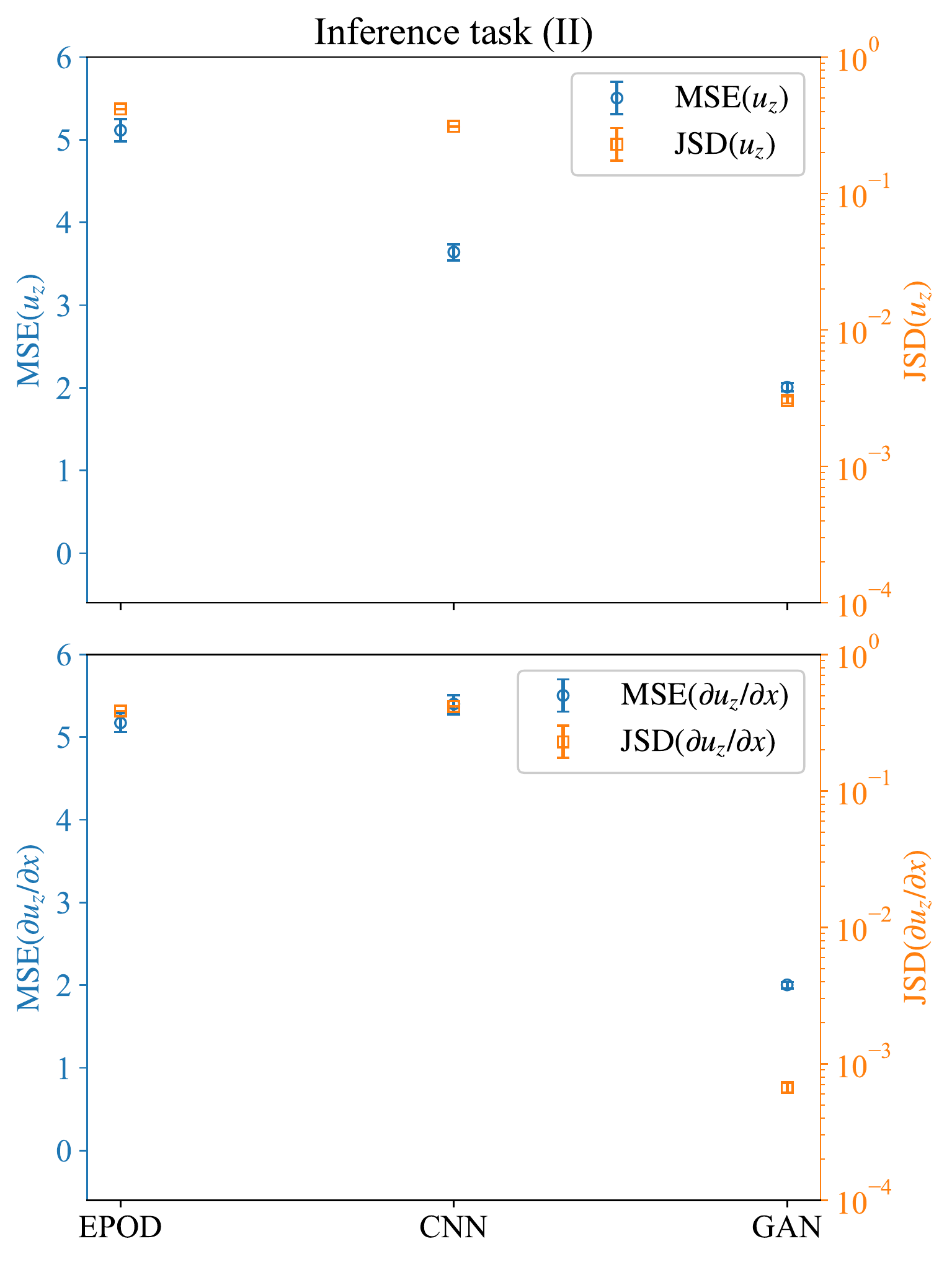}
	\caption{MSE (left y-axis) and JS divergence (right y-axis)between PDFs for the velocity component to be inferred, $u_z$, and its gradient, $\pt u_z/\pt x$. Results are obtained from the different tools for the inference task (II).}
	\label{fig:MSE-JSD-uz-ux}
\end{figure}

\begin{figure}
	\centering
	\includegraphics[width=1.0\linewidth]{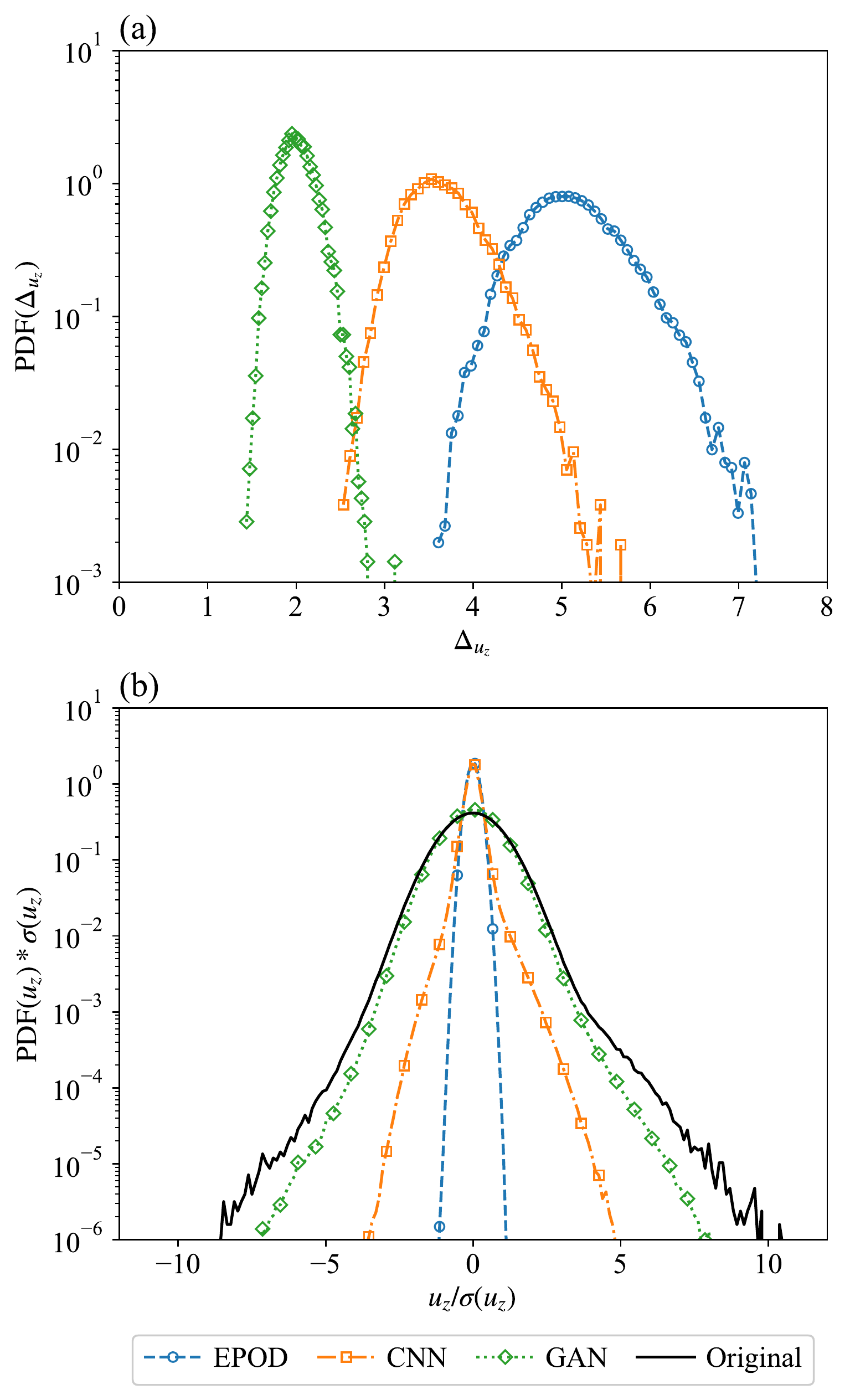}
	\caption{(a) PDFs of the spatially averaged $L_2$ error over different configurations and (b) PDFs of the predicted and the original velocity components, $u_z$, where $\sigma(u_z)$ is the standard deviation of the original data. Results are obtained from EPOD, CNN and GAN for the inference task (II).}
	\label{fig:PDF_L2_c-PDF-uz-ux}
\end{figure}

Fig. \ref{fig:MSE-JSD-uz-ux} shows MSEs of the interested component, $u_z$, and its gradient, $\pt u_z/\pt x$.
%\begin{table*}[h]
%	\begin{center}
%		\begin{minipage}{\textwidth}
%			\caption{MSE and JS divergence between PDFs for the velocity component to be inferred, $u_z$, and its gradient, $\pt u_z/\pt x$. Results are obtained from the different tools for the inference task (ii).}\label{tab:task2}%
%			\begin{tabular}{@{}cccccc@{}}
%				\toprule
%				& GPOD (DR) & GPOD (Lasso) & EPOD & CNN & GAN \\
%				\midrule
%				$\mathrm{MSE}(u_z)$ & $16.91_{-0.43}^{+0.48}$ & $2.605_{-0.058}^{+0.059}$ & $3.718_{-0.088}^{+0.095}$ & $3.490_{-0.105}^{+0.109}$ & $2.002_{-0.048}^{+0.040}$ \\
%				$\mathrm{JSD}(u_z)$ & $7.00_{-0.02}^{+0.03}\times10^{-1}$ & $1.43_{-0.02}^{+0.04}\times10^{-1}$ & $3.01_{-0.01}^{+0.01}\times10^{-1}$ & $2.94_{-0.01}^{+0.02}\times10^{-1}$ & $3.20_{-0.17}^{+0.16}\times10^{-3}$ \\
%				$\mathrm{MSE}(\pt u_z/\pt x)$ & $22.31_{-0.48}^{+0.47}$ & $2.212_{-0.040}^{+0.041}$ & $3.749_{-0.077}^{+0.079}$ & $5.082_{-0.118}^{+0.104}$ & $2.004_{-0.038}^{+0.035}$ \\
%				$\mathrm{JSD}(\partial u_z/\partial x)$ & $7.17_{-0.01}^{+0.02}\times10^{-1}$ & $4.91_{-0.14}^{+0.21}\times10^{-2}$ & $2.71_{-0.00}^{+0.00}\times10^{-1}$ & $3.98_{-0.01}^{+0.01}\times10^{-1}$ & $8.22_{-0.55}^{+0.37}\times10^{-4}$ \\
%				\botrule
%			\end{tabular}
%		\end{minipage}
%	\end{center}
%\end{table*}
Compared with the inference task (I), MSEs have larger values for the different tools, which indicates the difficulty of the task. For both $\mathrm{MSE}(u_z)$ and $\mathrm{MSE}(\pt u_z/\pt x)$, EPOD and CNN give large values, while the best values are around 2 from the GAN prediction. The PDF of the spatially averaged $L_2$ error, $\Delta_{u_z}$, over flow configurations are shown in Fig. \ref{fig:PDF_L2_c-PDF-uz-ux}(a). It shows that GAN has a peak of PDF with the smallest $\Delta_{u_z}$ over CNN and EPOD, which is consistent with Fig. \ref{fig:MSE-JSD-uz-ux}(top).
%Note that in the Appendix D of a previous work \cite{li2022data} we showed that uncorrelated predictions with correct statistics can lead to a MSE of 2. An example is using the randomly picked ground truth as the prediction. To check the correlation between the inferred and the original quantities, we show the ground truth and predictions of velocity component $u_z$ in 
We also present the JS divergence of $u_z$ and $\pt u_z/\pt x$ in Fig. \ref{fig:MSE-JSD-uz-ux}. Only GAN gives small values of JS divergence with the turbulent predictions. Fig. \ref{fig:PDF_L2_c-PDF-uz-ux}(b) further shows PDFs of $u_z$ from predictions, where GAN has the closest PDF with the original data.

\begin{figure}
	\centering
	\includegraphics[width=1.0\linewidth]{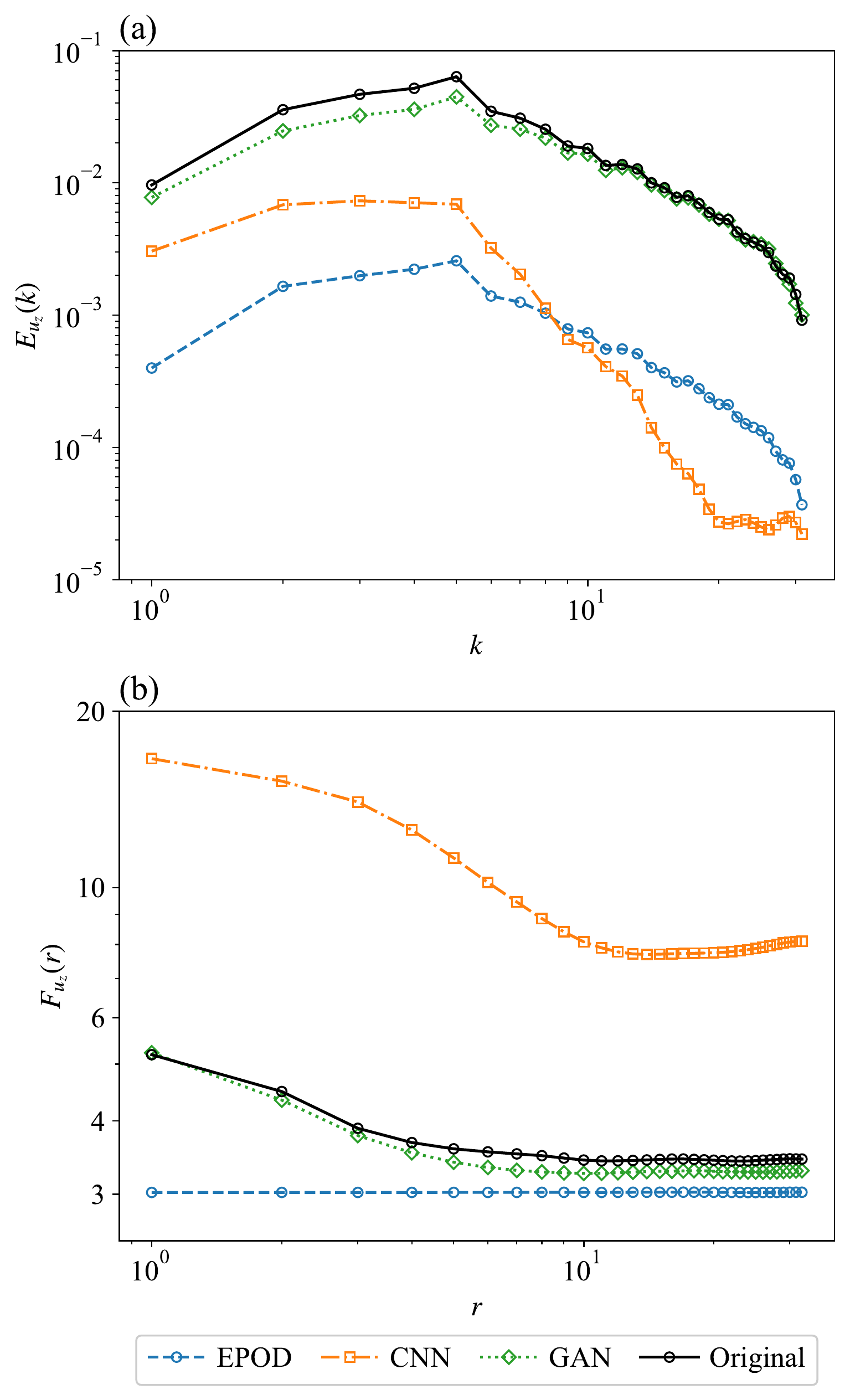}
	\caption{(a) The energy spectrum and (b) the flatness of predictions and the ground truth of the velocity component, $u_z$. The unit of $r$ is the grid width, $w_g=2\pi/64$. Results are obtained from the different tools for the inference task (II).}
	\label{fig:Spectrum-Flatness-uz-ux}
\end{figure}

To quantitatively study the predicted statistical properties, in Fig. \ref{fig:Spectrum-Flatness-uz-ux} we present the energy spectrum $E_{u_z}(k)$ and the flatness $F_{u_z}(r)$ for the different inferred fields.
%Since $u_z$ has a zero mean value, $E_{u_z}(k)$ should be negligible at $k=0$. Only EPOD satisfies this, but the other tools still predict small values of $E_{u_z}(k=0)$. However, at the other wave numbers only 
Only GAN predicts close energy to that of the original data at all wave numbers, which indicates that GAN generates inference with satisfying multi-scale properties (Fig. \ref{fig:Spectrum-Flatness-uz-ux}(a)). Fig. \ref{fig:Spectrum-Flatness-uz-ux}(b) shows that the flatness from GAN has a similar shape with the original one. They are close at small and intermediate scales but that of GAN has smaller values at large scales. Besides, all the other tools cannot predict satisfying flatness at all scales.

\section{Conclusion}\label{sec:Conclusion}

We studied a problem with practical applications in geophysical and engineering problems, which is using one velocity component to infer another one for 2D snapshots of rotating turbulent flows. Moreover, this problem is also with theoretical interest, connected to the feature ranking of turbulence: identifying which degrees of freedoms are dominant in turbulent flows.

Linear (EPOD) and non-linear (CNN \& GAN)  methods are systematically compared by two inference tasks with different complexities, which depend on the correlation between the input component and the one to be inferred. The EPOD method conducts POD analysis with the supplied component and gives the correlated part as the prediction. CNN \& GAN methods are based on fully non-linear approximators, without and with an adversarial component.

For the simpler inference task (I), EPOD generates meaningful inference which is not satisfying in terms of the MSE and the JS divergence with real data. An obvious improvement is reached after the usage of non-linear mapping, where CNN gives good predictions with small MSE and JS divergence with the ground truth. Compared with CNN, GAN makes the inference more realistic with fine structures, with the cost of slightly increasing the MSE. Besides, the training cost is also more expensive for GAN. For the inference task (II), the supplied component is not well correlated with the one to be inferred. The EPOD method cannot make meaningful inference because of this complexity. However, with the non-linear capability CNN and GAN can recognize the positions of coherent structures, although they cannot predict correct values (or even signs) in the vortices. Moreover, the usage of discriminator is very important for this task. Indeed,  CNN only predicts some smooth blobs while GAN generates turbulent configurations correlated with the ground truth. 
In conclusion, we have shown that GANs are simultaneously optimizing the $L_2$ point-to-point distance, snapshot by snapshot, and the cross entropy between the ground-truth data and the generated ones, leading to both instantaneous and statistical reconstruction optima. On the other hand, EPOD minimizes only the variance through the eigenvalues of each single field. The latter yields a worse result of statistical properties.
%\blue{In conclusion, we have shown that  GANs are simultaneously  optimising  the cross entropy between the ground-truth data and the generated ones, and the $L_2$ point-to-point distance, image-by-image, leading to  both instantaneous and statistical reconstruction optima.  On the other hand,  EPODs do minimise only the variance through the eigenvalues of each single image. The latter yields a worse result for probability.}

%Finally, we also showed that GAN predictions always have good statistical properties.

Note that EPOD, CNN and GAN-based methods can  se straightforwardly generalized  to the inference task with multiple components/quantities supplied and/or to be inferred. The inference task where
\begin{equation}
	\bm{u}_S\colon(u_x,u_y)\to\bm{u}_G\colon u_z \nonumber
\end{equation}
was also conducted and it shows very similar results with  task (II), because of the high correlation between the two in-plane components. Future research can give more consideration to serving practical applications such as PIV. As the current research focuses on fully resolved turbulent flow, future studies can focus on conducting inference with noisy and/or Fourier-filtered data. Additionally, it is recommended to investigate the potential benefits of employing machine learning techniques \cite{hochreiter1997long, sutskever2014sequence, vaswani2017attention} to exploit the temporal correlations present in the data. Given that the current study is limited to 2D fields, it is impossible to impose physical constraints such as incompressibility. A possible avenue for future research is to consider fields in a 3D volume, which may better accommodate the imposition of such prior knowledge.

\medskip

\noindent\textbf{Acknowledgments.}
This work was supported by the European Research Council (ERC) under the European Union’s Horizon 2020 research and innovation programme (Grant Agreement No. 882340)

\section*{Declarations}
\textbf{Data availability statement.}
The datasets that support the findings of this study are openly available in the Smart-Turb repository \cite{biferale2020turb} at \url{http://smart-turb.roma2.infn.it}.

\medskip

\noindent\textbf{Author contributions.}
All authors contributed equally to the paper.

\begin{appendices}

\section{}\label{secA1}

We choose the parameter $\alpha=0.2$ for all leaky ReLU activation functions. Adam optimizer \cite{kingma2014adam} is used to train the generator and the discriminator simultaneously, where the learning rate of generator is twice as that of the discriminator. A staircase-decay schedule is adopted to the learning rate in order to improve the stability of training, where the learning rate decays with a rate of 0.5 every 50 epochs for 11 times, corresponding to the maximum epoch of 600. The initial learning rate of the generator is $10^{-3}$ and we use a batch size of 128. Results show that $\mathcal{L}_\mathrm{MSE}$ (Eq. (\ref{eq:L_MSE})) and $\mathcal{L}_{adv}$ (Eq. (\ref{eq:L_adv})) fluctuate around fixed values at later epochs, which indicates the convergence of training.

\begin{figure}
	\centering
	\includegraphics[width=1.0\linewidth]{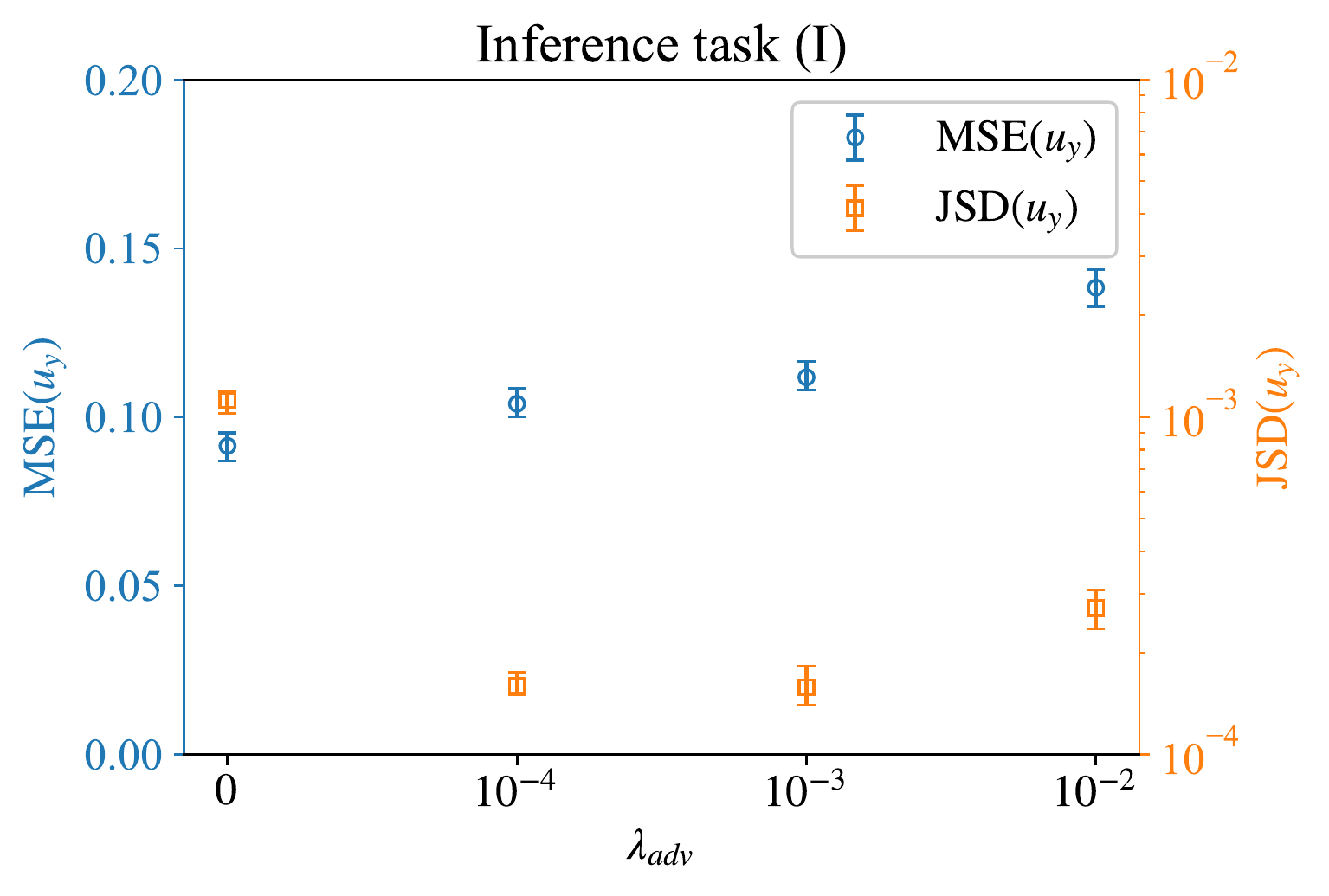}
	\caption{MSE and JS divergence between PDFs for the velocity component to be inferred, $u_y$. Results are obtained from GANs with different adversarial ratios, $\lambda_{adv}$, for the inference task (I). }
	\label{fig:MSE-JSD-ARs-uy-ux}
\end{figure}

Fig. \ref{fig:MSE-JSD-ARs-uy-ux} presents the MSE and JS divergence between PDFs of $u_y$ from GANs with different adversarial ratios, $\lambda_{adv}$, for the inference task (I). It shows that a non-zero $\lambda_{adv}$ can reduce the JS divergence with enlarging the MSE slightly. We choose $\lambda_{adv}=10^{-3}$ for all the GANs in this study.




\end{appendices}


%\bibliographystyle{bst/sn-basic}
\bibliographystyle{unsrt}
\bibliography{sn-bibliography}% common bib file


\end{document}